\documentstyle[12pt]{article}
\begin{document}
\begin{center}

{\large \bf Analog of the Fizeau Effect in an Effective Optical Medium}\\

\vspace{8mm}
 K.K. Nandi,$^{~ a,c,}$\footnote{E-mail address: kamalnandi@hotmail.com}
  Yuan-Zhong Zhang,$^{~ b,c,}$\footnote{E-mail address: yzhang@itp.ac.cn}
  P.M.Alsing,$^{~ d,}$\footnote{E-mail address: alsing@ahpcc.unm.edu}
  J.C.Evans$^{~ e,}$\footnote{E-mail address: jcevans@ups.edu}
and A.Bhadra$^{~ f,}$\footnote{E-mail address:
            aru\_bhadra@yahoo.com} \\
\vspace{4mm}
  {\footnotesize
  {\it
  $^a$Department of Mathematics, University of North Bengal,
      Darjeeling WB 734430, India\\
  $^b$CCAST(World Lab.), P. O. Box 8730, Beijing 100080, China \\
  $^c$Institute of Theoretical Physics, Chinese Academy of Sciences,
          P.O. Box 2735,\\ Beijing 100080, China\\
  $^d$Albuquerque High Performance Computing Center, University of New Mexico,
      Albuquerque, New Mexico, NM 87131\\
  $^e$Department of Physics, University of Puget Sound, Tacoma, Washington,
      WA 98416\\
  $^f$High Energy and Cosmic Ray Research Center, University of North Bengal,
      Darjeeling, WB 734430, India.}}
\end{center}

\vspace{8mm}

\begin{abstract}
Using a new approach, we propose an analog of the Fizeau effect
for massive and massless particles in an effective optical medium
derived from the static, spherically symmetric gravitational
field. The medium is naturally perceived as a dispersive medium by
matter de Broglie waves. Several Fresnel drag coefficients are
worked out, with appropriate interpretations of the wavelengths
used. In two cases, it turns out that the coefficients become
independent of the wavelength even if the equivalent medium itself
is dispersive. A few conceptual issues are also addressed in the
process of derivation. It is shown that some of our results
complement recent works dealing with real fluid or
optical black holes.\\

\noindent
  PACS number(s): 03.65.Bz, 42.15.- i, 04.20.-q
\end{abstract}

\vspace{10mm}
 \section{Introduction and reappraisals}
The historical Fizeau effect for light in moving media has been
reconsidered by several authors [1-5] in recent times. We shall
consider it here in the context of static, isotropic gravity. To
our knowledge, such an investigation has not been undertaken
before. We deemed it worthwhile to examine how an old effect would
look like in a new theoretical model and what conceptual issues
are involved. However, we must make clear at the outset that the
{\it only} quantity to be borrowed from general relativity is the
effective refractive index. The rest of the analysis is special
relativistic (see Sec. 3).

In the literature, generally, the Fizeau effect is considered in
connection with its close relative, the analog of the
Aharonov-Bohm (AB) effect in real material medium. Several
interesting results have followed from these analyses. For
instance, Leonhardt and Piwnicki [6] have shown that a
nonuniformly moving medium appears to light as an effective
gravitational field for which the curvature scalar is nonzero.
They also show how light propagation at large distances around a
vortex core shows Aharonov-Bohm (AB) effect and at shorter
distances resemble propagation around what are termed as optical
black holes. Berry {\it et al} [7] demonstrated the AB effect with
water waves and Roux {\it et al} [8] observed it for acoustical
waves in classical media. The curved space analogy has been
predicted for fluids and superfluids as well [9]. The spirit of
the present work, in some sense, is in a direction that is reverse
to the idea of the above curved space analogy. That is, our
interest is to calculate the Fizeau type effect for {\it both}
massless and massive particles in a static, spherically symmetric
gravity field but portraying it as an effective optical medium. In
the process, we shall also see the extent to which the curved
space analogy compare with the results derived in a genuine
gravitational field. For simplicity, we shall assume only uniform
motion of our effective medium resulting from the relative motion
between the gravitating source and the observer.

An outline of the Fizeau effect is this: Consider a tube through
which a fluid with a refractive index $n$ is flowing with velocity
$V$. Then, let light pass through the tube parallel to its axis.
In the comoving frame of the water, the speed of light is
$v'(=c_0/n)$, but in the frame in which the water appears to be
flowing, the speed of light has been found to be:
  $$v = v' \pm \left( {1 - \frac{{v'^2 }}{{c_0^2 }}} \right)V
  + O(V^2)\approx \frac{{c_0 }}{n}\pm \left({1-\frac{1}{{n^2 }}}
  \right)V,                                     \eqno(1)$$
where $c_0$ is the speed of light in vacuum. The quantity
$(1-n^{-2})$ is called the Fresnel drag as he was the first to
predict it theoretically. Obviously, the resultant speed $v$ does
not conform to the Galilean law of addition of velocities $v'\pm
V$. The effect, after it was experimentally observed by Fizeau in
1851, was regarded as an empirical fact awaiting a correct
theoretical interpretation. It came only after the advent of
Einstein's special theory of relativity in 1905. It has since been
realized that the Fizeau effect symbolizes only a first order
approximation of the exact one dimensional special relativistic
velocity addition law (VAL) derived from Lorentz spacetime
transformations. Originally, Fizeau did not consider dispersion
but nowadays it is recognized that the effect also contains a term
due to the effect of dispersion.

In our investigation, we shall adopt an approach involving quantum
mechanics, general and special relativity using the method of what
is known as the optical-mechanical analogy. The historical and
fundamental role of the analogy in the development of modern
theoretical physics need not be emphasized. Apart from the crucial
role it played in the development of quantum mechanics, especially
in the de Broglie wave-particle duality, it provides an excellent
tool that enables one to visualize the problems of geometrical
optics as problems of classical mechanics and vice versa.

In a series of papers [10], it has been shown that the
optical-mechanical analogy can be recast into a familiar form that
allows one to envisage the mechanical particle equation as a
geometrical optical ray equation and the latter as a Newtonian
``$F=ma$" equation:
 $$\frac{{d^2 \vec r}}{{dA^2 }} = \vec \nabla \left( {\frac{1}{2}n^2
 c_0^2 } \right),                                  \eqno(2)$$
 $$dA = \frac{{dt}}{{n^2 }},                       \eqno(3)$$
where $\vec r \equiv (x,y,z)$ or $(r,\theta ,\varphi )$ ,
$\vec\nabla$ is the gradient operator, $n$ is the index of
refraction, not necessarily constant, and $A$ was originally
called the stepping parameter but it could also be identified as
the optical action and related to several other physical
parameters. Many illustrations in ordinary gradient index optics
demonstrated the validity of the Eqs.(2) and (3) and their
usefulness as a heuristic tool.

An interesting turn in the direction of investigation is signaled
by the introduction of general relativity [11]. Exact equations
for light propagation in the static, spherically symmetric field
of Schwarzschild gravity do indeed follow from the Eqs.(2) and (3)
when an appropriate gravitational index of refraction $n(\vec r)$
is employed. The analysis also brings forth the distinct but
complementary roles played by the optical action $A$ and
coordinate time $t$. To see this, note that the first integral of
Eq.(2) is
 $$\left| {\frac{{d\vec r}}{{dA}}} \right| = nc_0,  \eqno(4)$$
or equivalently, using Eq.(3),
 $$\left| {\frac{{d\vec r}}{{dt}}} \right| = \frac{{c_0 }}{n}.
                                                    \eqno(5)$$
However, the force laws have changed thereby. In Eq.(4), the
"potential" is $\frac{1}{2}n^2 c_0^2$ while in Eq.(5), the
potential is $ - \frac{1}{2}\frac{{c_0^2 }}{{n^2 }}$. On
eliminating $A$ from Eq.(4) or $t$ from Eq.(5), we would therefore
obtain two path equations for light on a plane, but {\it only} the
former, not the latter, gives the right answer. On the other hand,
Eq.(5) gives the correct equation for the Shapiro time delay
$\Delta t$, while $\Delta A$ from Eq.(4) does not. A deeper
understanding of the parameter $A$ is still awaited.

Our basic strategy is to regard the gravity field as an effective
refractive optical medium imposed on a {\it fictitious} Minkowski
space so that Lorentz transformations can be used to relate two
relatively moving observers in that space. (Note that we are not
talking of a division of the metric tensor into two parts, but
rather of a scalar field placed upon a flat space. For more
discussion, see Sec. 3). This is just an intermediate exercise.
The final outcome has to be translated back into the actually
observable quantities in a gravity field. The idea that a gravity
field could be formally equivalent to a refractive medium with
respect to optical propagation is not new. It goes all the way
back to Eddington [12] who was the first to advance the expression
of a gravitational refractive index in an approximate form. It was
used later, in varying degrees, by several other researchers
[13,14] in the investigation of specific problems. But none of the
works really focused on how the {\it exact} general relativistic
equations of trajectories, frequency shifts or Shapiro time delay
for massless particles could be obtained in that equivalent
medium. The motion of massive particles was not addressed at all.
The extension of the work in Ref.[11] that includes also the
massive particle motion now exists [15,16]: A suitably modified
index of refraction together with the ``$F=ma$" formulation
immediately reproduce all the desired exact equations in the
static, isotropic gravity field. The method has been applied very
successfully to Friedmann cosmologies as well that yielded some
new interesting insights. All the above systematic developments
amply indicate the usefulness of the concept of an effective
gravitational index of refraction. By way of a further extension,
the index has been calculated also for a more general class of
rotating metrics [17]. A new and significant development has come
in the shape of a most recent formulation [18] of a single set of
unified optical-mechanical equations that allow easy introduction
of quantum relations into it. As a consequence, one then finds
that massive de Broglie waves necessarily perceive the gravity
field as a {\it dispersive} optical medium.

In this paper, our basic aim hinges around calculating the
consequences arising out of this dispersion in the form of what
may be termed as the gravitational Fresnel drag, {\it dispersion
included}. There are several spin offs: It will be demonstrated
that, in the comoving frame, the expressions for the Lagrangian
and the dispersion relation are similar to those obtained by
Leonhardt and Piwnicki [6] in the context of real media. These
similarities provide a direct extension of these expressions in a
realistic gravity field. It will also be evident that the
conditions for optical black holes [6,19,20] are naturally met in
the equivalent medium, irrespective of whether one considers light
or massive de Broglie waves.

The paper is organized as follows: Sec. 2 contains a brief survey
of the basic equations that will be used throughout the paper.
Conceptual justifications for the adopted procedure appear in Sec.
3 . In Sections 4--6, the gravitational Fresnel drags are
calculated for different choices of the wavelengths. Sec. 7
contains a brief discussion of operational definitions. In Sec. 8,
we demonstrate how the results dealing with a real fluid medium
compare with those in a genuine gravity field considered in this
paper. Finally, in Sec. 9, we summarize and add some
remarks.\\

\section{Basic equations}

Consider a static, spherically symmetric, but not necessarily
vacuum, solution of general relativity written in isotropic
coordinates
 $$ds^2  = \Omega ^2 (\vec r')c_0^2 dt'^2  - \Phi ^{ - 2} (\vec
  r')\left| {d\vec r'} \right|^2,            \eqno(6)$$
where $\Omega$ and $\Phi$ are the solutions of Einstein's field
equations. Many metrics of physical interest can be put into this
isotropic form including the experimentally verified Schwarzschild
metric. The coordinate speed of light $c(\vec r')$ is determined
by the condition that the geodesic be null ($ds^2 =0$ ):
 $$c(\vec r') = \left| {\frac{{d\vec r'}}{{dt}}} \right| = c_0 \Phi
(\vec r')\Omega (\vec r').                      \eqno(7)$$
 We take leave from the metric approach at this point and define the
effective index of refraction for light in the gravitational field
as
 $$n(\vec r') = \Phi ^{ - 1} \Omega ^{ - 1}.     \eqno(8)$$

We shall omit further details here that can be found in Ref.[18],
but only state the results to be used in this paper. The first
step in the direction of introducing quantum mechanics in a
semiclassical way is to have a single refractive index $N$ and a
single set of equations that should be valid for both massless and
massive particles. The result is:
 $$\frac{{d^2 \vec r'}}{{dA^2 }}=\vec \nabla \left( {\frac{1}{2}N^2
   c_0^2 } \right), ~~~~ {\rm ( light\, and\, particles)} \eqno(9)$$
 $$\left| {\frac{{d\vec r'}}{{dA}}} \right| = Nc_0,
     ~~~~ {\rm ( light\, and  \, particles)}             \eqno(10)$$
where, once again, it is the same $A$, satisfying $dA =
\frac{{dt}}{{n^2 }}$, that appears even for massive particle
trajectories. It looks as if the action has a foot in the wave
regime and a foot in the particle regime. The second step involves
the introduction of the Planck relation $H' = \hbar \omega '$ and
the de Broglie relation $p' = \hbar k' = h/\lambda '$, where $h =
2\pi \hbar$, in the expression for $N$. As usual, $H'$ and $p'$
are the total energy and momentum respectively, $\omega '( \equiv
2\pi \nu ')$  and $\lambda '$ are the coordinate frequency and the
wavelength of the de Broglie waves. The physically measurable
corresponding proper quantities are $\tilde \omega ' = \omega
'/\Omega $ and $\tilde \lambda ' = \lambda '/\Phi $ respectively.
The third step finally gives the desired index of refraction $N$
of the dispersive medium due to massive de Broglie waves:
 $$N(\vec r',\omega ') = n(\vec r')\sqrt {1 - \frac{{m^2 c_0^4 \Omega
  ^2 (\vec r')}}{{\hbar ^2 \omega '^2 }}},           \eqno(11)$$
where $m$ is the rest mass of the test particle. One may also
rewrite $N$ as
 $$N=\frac{{c_0 p'}}{{H'}}=\frac{{n^2 v'}}{{c_0 }},  \eqno(12)$$
where $v'$ is the (unobservable) coordinate speed of the classical
particle in the medium. Also it follows that
 $$\omega ' =2\pi\nu ',\quad \;\lambda ' =\frac{{c_0 }}{{N\nu '}}.
                                                     \eqno(13)$$
Using Eq.(13), $N$ can be rewritten in a more transparent form:
 $$N =\frac{{n(\vec r')}}{{\sqrt{1+\left({\frac{{\tilde\lambda'}}
     {{\lambda _c }}} \right)^2 } }},                \eqno(14)$$
where $\lambda_c =h/mc_0$  is the Compton wavelength of the
particle. Clearly, for light, $m=0, N=n$,   and one recovers
Eqs.(2) and (3) from Eqs.(9) and (10) respectively. That is, light
waves do not perceive the effective medium as dispersive. However,
for $m\ne 0$, dispersion seems inevitable, as evidenced from
Eqs.(11) or (14), if quantum relations are introduced.

We shall require also the following: The mass shell constraint is
given by [18]:
 $$\hbar ^2 \omega '^2 = m^2 c_0^4 \Omega ^2  +\frac{{c_0^2 \hbar
  ^2 k'^2 }}{{n^2 }}.                         \eqno(15)$$
The phase velocity is
  $$v'_p =\frac{{H'}}{{p'}}=\frac{{\omega '}}{{k'}} =\frac{{c_0
  }}{N},\quad \quad v'_p v'_g  = \frac{{c_0^2 }}{{n^2 }},
                                               \eqno(16)$$
giving the group velocity
 $$v'_g  = \frac{{d\omega '}}{{dk'}}=\frac{{c_0 N}}{{n^2 }}= v'.
                                              \eqno(17)$$
It should be mentioned that the validity of the expression (11) is
established also by the WKB analysis of the massive generally
covariant Klein-Gordon equation [18]. Moreover, the mass shell
constraint (15) yields the {\it exact} Stodolsky phase [21] in the
case of spin-1/2  Dirac equation in curved spacetime [22]. This
last result is extremely interesting.

With the Eqs.(6)-(17) at hand, we are able to calculate the
Fresnel drag factors under different scenarios, but, before this,
we need to clear up a few relevant concepts. Note that all the
expressions in this section refer to the comoving frame, that is,
the frame fixed to the gravitating source. Henceforth, in order to
have conformity with notations in the literature, all expressions
in the comoving frame will be designated by primes and those in
the relatively moving lab frame will be denoted by unprimed
ones.\\

\section{Conceptual issues}

The following discussion is aimed at providing appropriate
interpretations of the quantities that appear in the various
formulations of the Fizeau effect. There are two basic
ingredients. The first is the VAL. In many works dealing with the
effect, the one dimensional VAL, which is valid for point
particles, is employed, implicitly or explicitly, also for {\it
waves} propagating with the phase speed $c_0 /n$. The procedure is
to use the one dimensional Lorentz transformation equations in the
form
 $$\omega ' = \gamma (\omega  - kV),\quad \quad k' = \gamma (k -
  V\omega c_0^{ - 2} ),                       \eqno(18)$$
 $$\omega  = \gamma (\omega ' + k'V),\quad \quad k = \gamma (k' +
   V\omega 'c_0^{ - 2} ), \quad \quad
   \gamma  = (1 - V^2 /c_0^2 )^{- 1/2}        \eqno(19)$$
and obtain a VAL as
 $$v_p ' = (v_p  - V)(1 - v_p V/c_0^2 ), \quad \quad
 v_p  = (v'_p  + V)(1 + v'_p V/c_0^2 ),        \eqno(20)$$
where $v'_p  = \omega '/k' = c_0 /n$ is the phase speed of light
in the (primed) comoving frame of the medium and $v_p =\omega /k$
is the phase speed in the (unprimed) lab frame in which the medium
appears to be moving with uniform relative velocity $V$. The phase
speed however could well exceed $c_0$ in many physical
configurations where $n<1$.

On the other hand, an {\it a priori} prescription that $n$ be
greater than unity (making $c_0 /n < c_0$) somewhat diminishes the
generality of the theory. However, this deficiency may not pose
any realistic problem in a nondispersive medium. When dispersion
is involved, the most appropriate quantity to use in the VAL is
the group speed $d\omega /dk$ (that involves the knowledge of
$dn/d\omega$), which simply equals the velocity of the classical
point particle, rather than the phase speed. As stated before, the
original Fizeau experiment did not consider any dispersion, the
index $n$ was taken to be a true constant, so that the group and
the phase velocities coincided precisely to $c_0 /n$. In general,
they are different for massive de Broglie waves, as our later
equations will reveal. In our calculation of the Fizeau effect,
the mass shell constraint, Eq.(15), or, by another name, the
dispersion relation plays a key role: It provides well defined
expressions for the group and phase velocities. Such types of
natural constraints are unavailable in just any arbitrary medium
consisting of solids or liquids. In these cases, dispersion is
normally introduced by hand.

An important point should be noted here. In describing the Fizeau
experiment with ordinary medium (such as water), one takes the
background spacetime to be flat. Such Minkowski networks, composed
of rods and clocks, are actually unobservable in a gravity field
due to the universality of gravitational interaction, or, saying
more technically, due to the principle of equivalence. There does
not exist a unique division of the metric tensor into a background
and a field part. We consider here a different kind of separation
according to which the gravity field is looked upon as analogous
to an optical medium imposed upon a flat background spacetime, the
index $N$ summarizing the nonlinearities of the gravity field, as
it were. The important point is that the analogy, though intended
to be only of {\it formal} nature, may lead to results that could
be testable by experiment (see Sec. 9 for a discussion). With this
understanding, let us conceive of observers equipped with
fictitious Minkowski networks and apply, as an intermediate step,
the full machinery of special relativity in what follows.

Thus, we take Eq.(19) in the form
 $$\Delta \omega  = \gamma (\Delta \omega ' + \Delta k'V),\quad \quad
  \Delta k = \gamma (\Delta k' + V\Delta \omega 'c_0^{ - 2} ),
                                                  \eqno(21)$$
which give the VAL, denoting $v_g '=\frac{{\Delta \omega '}}
{{\Delta k'}}$, as:
 $$v_g  = \frac{{\Delta \omega }}{{\Delta k}}=\frac{{v_g '+V}}{{1
+ Vv'_g /c_0^2 }}.                                \eqno(22)$$

The second ingredient is the special relativistic Doppler shift in
one dimension giving frequency (or wavelength) transformation
between two frames in relative motion. Thus, one takes Eq.(19) in
the form
 $$\omega  = \gamma \omega '(1 + k'V/\omega '),   \eqno(23)$$
and specifies $\omega '/k'$. At this point, let us note that,
Cook, Fearn and Milonni [2] have considered two possibilities in
the context of Fizeau experiment with real media having refractive
indices $n$.\\

{\underline {\bf Case 1}}

Take $\omega '/k'=c_0$  in Eq.(23). This case has been considered
by Synge [3]. That is, take the usual Doppler shift formula,
which, written in terms of the wavelength, is
 $$\lambda  = \lambda '\sqrt {\frac{{1 - \frac{V}{{c_0 }}}}{{1 +
  \frac{V}{{c_0 }}}}}.                       \eqno(24)$$
The corresponding physical configuration consists of a block of
material moving with velocity $V$ in an otherwise empty lab frame.
The wavelength $\lambda'$ of a light pulse measured by an observer
stationed at the interface between the block and the empty space
will appear to the lab observer as $\lambda$ according to Eq.(24).
Inside the block, however, $\lambda'$ is assumed to be a constant.
The resulting Fresnel drag has been experimentally confirmed to a
very good accuracy by Sanders and Ezekiel [4].

Unfortunately, it is difficult to conceive of a parallel
configuration in our problem. The entire optical medium can
neither be simply put inside a box with a certain boundary nor
need the wavelength $\lambda'$ be constant throughout the medium.
Instead, it is easier to consider two relatively moving observers
associated with the background empty frame who may use Eq.(24). We
have to calculate how one observer translates the observations of
another at a certain point when they happen to pass each other.
This is done in Sec. 4.\\

{\underline{ \bf Case 2}}

Take $\omega'/k'=c_0/n$ in the Eq.(19) for $k$. Cook {\it et al}
[2] provide the corresponding physical configuration in this case.
According to Lerche [1], the lab observer can exercise two
options. Either he/she uses (i) a wavelength $\lambda$ given by
the Doppler formula (23) but with $\omega'/k'=c_0/n$ or uses (ii)
a vacuum wavelength $\lambda _0  = 2\pi c_0 /\omega$. The forms
for the drag coefficients will be different in the two cases. The
parallel options in our case are the same, except that we have to
use $N$ instead of $n$, so that the Doppler formula reads
$$\lambda  = \frac{{\lambda '}}{\gamma }\left( {1 + \frac{V}{{Nc_0
  }}} \right)^{ - 1}.                           \eqno(25)$$
We shall work out both the options in Sec. 5. This particular
formula appears to be more consistent with our formulation {\it
per se} as we will be using our own definition of $\omega '/k'$
given in Eq.(27). We can also add a {\it third} possibility worked
out in Sec. 6. This is a special feature of the gravitational case
we are considering.\\

{\bf \underline{ Case 3}}

Consider a stationary observer $\tilde A$ at a point in the
gravity field measuring proper (or physical) wavelength
$\tilde{\lambda'}$. He/she also measures the proper velocity of
light in his/her neighborhood to be just $c_0$. A freely falling
observer $\tilde B$ at that point, having an instantaneous
velocity $\tilde V$ relative to $\tilde A$, would measure
$\tilde{\lambda}$ according to the options, which, to first order,
are
$$\tilde \lambda ' \approx \tilde \lambda (1 + \tilde V/c_0 ),
\quad \quad \tilde \lambda ' \approx \tilde \lambda (1 + \tilde
V\tilde N/c_0 )\quad \quad \tilde \lambda _0  = 2\pi c_0 /\tilde
\omega .                                           \eqno(26)$$
Note that there is a difference between the present stationary
observer and a stationary observer associated with the background
flat space of Case 1: The group velocities of the matter de
Broglie waves measured by them are not the same (see below). We
now proceed to calculate the Fresnel drags successively in all the
three cases using the same VAL, Eq.(22), but different Doppler
formulas, Eqs.(24)--(26).\\

\section{ Fresnel drag: Case 1}
Suppose that an observer A, equipped with a Minkowski network, is
at rest at $r=0$ in a spherically symmetric medium. He/she
measures the coordinate phase and group velocities respectively of
a massive de Broglie wave packet at $r=r_{0}'$ as, using Eqs.(15):
$$\frac{{\omega'}}{{k'}}=v'_p =\frac{{c_0 }}{{N(r'_0 ,\lambda')}},
                                                \eqno(27)$$
$$\frac{{\Delta \omega '}}{{\Delta k'}} = v'_g  = \frac{{c_0
}}{{\bar N(r'_0 ,\lambda ')}} < c_0 ,\quad \quad \bar N =
\frac{{n^2 }}{N} > 1.                           \eqno(28)$$
 For a light pulse, $v'_p  = v'_g  = \frac{{c_0 }}{n}$ and these
are independent of the wavelength $\lambda'$ or wave number $k'$.
The same holds for $v_g$ as well. This implies that the
trajectories of light rays do not depend on the wave properties of
light. However, in general, $v'_p  \ne v'_g$, as is evident from
Eqs.(27) and (28).

Consider another observer B moving in the same radial direction
with uniform velocity $V$ with respect to A. Then, in the frame of
B, identified as the lab observer, the entire medium moves
uniformly, that is, A becomes the comoving observer. How will B
translate the observations of A, when their origins coincide at
$r=0$? To find it out, note that the coordinate length $r_{0}'$
will appear to B as
$$r_0  = r'_0 \sqrt {1 - \frac{{V^2 }}{{c_0^2 }}}.\eqno(29)$$
Also, the velocity $v_{g}'$ observed by A will appear to B as
$v_g$ given by the special relativistic VAL, Eq.(22). We may
explicitly express $v_{g}'$ in terms of $(r_{0}',\lambda')$ as
$$v'_g  = \frac{{c_0 }}{{\bar N(r'_0 ,\lambda ')}} = \frac{{c_0
}}{{n(r'_0 ) \times \sqrt {1 + \left( {\frac{{\lambda '}}{{\lambda
_c }}} \right)^2 \Phi ^{ - 2} (r'_0 )} }}.         \eqno(30)$$
When this expression for $v_{g}'$ is plugged into the right hand
side of Eq.(22), one finds the answer to the question above:
$v_{g}(r_{0}',\lambda')$ is the exact radial group velocity of the
de Broglie waves to be observed by B. But B uses the Doppler
shifted wavelength $\lambda$ instead of $\lambda'$. Then, to first
order in $(V/c_{0})$ ), we get from Eq.(24):
 $$\lambda ' \approx \lambda (1 + V/c_0 )=\lambda + \Delta\lambda,
 \quad \quad r_0 ' \approx r_0 .                 \eqno(31)$$
Considering the right hand side of Eq.(30) and writing the
denominator as $\bar N(r_0 ' = r_0 ,\lambda ') \equiv \bar
N(\lambda  + \Delta \lambda )$, we get from the Taylor expansion
 $$\bar N(\lambda  + \Delta \lambda ) \approx \bar N(\lambda ) +
\Delta \lambda \frac{{\partial \bar N}}{{\partial \lambda }} =
\bar N(\lambda )(1 + \frac{{\lambda V}}{{c_0 \bar
N}}\frac{{\partial \bar N}}{{\partial \lambda }}).$$
 From Eqs. (22) and (28), we get, using the above, a redefined
index $\bar{N}'$ such that
  $$v_g (\lambda ') = \frac{{c_0 }}{{\bar N'(\lambda ')}} = \frac{{c_0
}}{{\bar N(\lambda ')}} + \left( {1 - \frac{1}{{\bar N^2 (\lambda
')}}} \right)V.                                    \eqno(32)$$ In
other words, in the approximation considered,
$\bar{N}^{2}(\lambda') \approx \bar{N}^{2}(\lambda)$ and we have,
 $$v_g (\lambda ) = \frac{{c_0 }}{{\bar N(\lambda  + \Delta \lambda
)}} + \left( {1 - \frac{1}{{\bar N^2 (\lambda )}}} \right)V =
\frac{{c_0 }}{{\bar N(\lambda )}} + F_1 V,          \eqno(33)$$
where
 $$F_1  \equiv \left( {1 - \frac{1}{{\bar N^2 (\lambda )}}} \right) -
\frac{\lambda }{{\bar N^2 (\lambda )}} \times \frac{{\partial \bar
N}}{{\partial \lambda }}                            \eqno(34)$$ is
the Fresnel drag we have been looking for. It can be easily
verified that the same $F_1$ follows also from the ordinary
expansion of $v_{g}\left(r'_{0},\lambda'\right)$ in Eq.(22) in
conjunction with Eqs.(21) and (30) under the small velocity
approximations, Eq.(31), but the steps as given above are the
simplest. For light waves, $\bar N \to n$, and one has
 $$F_1  \equiv \left( {1 - \frac{1}{{n^2 (\lambda )}}} \right) -
\frac{\lambda }{{n^2 (\lambda )}} \times \frac{{\partial
n}}{{\partial \lambda }}.                  \eqno(35)$$
Interestingly, although the dependence of $n$ on $\lambda$ is not
known, the dispersion nonetheless follows here as an inheritance
from Eq.(34). This is the formula proposed by Synge [3] and also
experimentally tested [4] with $n$ as the refractive index of the
block.

Using Eqs.(14) and (28), we can have the explicit expression from
Eq.(34) as:
  $$F_1  = 1 - \frac{1}{{n^2 (r_0 )\left[ {1 + \left( {\frac{{\tilde
\lambda }}{{\lambda _c }}} \right)^2 } \right]}} - \frac{{\left(
{\frac{{\tilde \lambda }}{{\lambda _c }}} \right)^2 }}{{n(r_0
)\left[ {1 + \left( {\frac{{\tilde \lambda }}{{\lambda _c }}}
\right)^2 } \right]^{3/2} }}.                       \eqno(36)$$

Note that, in the asymptotic region $r \to \infty$, or in the
absence of gravity, one has $n(r) \to 1,{\kern 1pt} \,\,\tilde
\lambda  \to \lambda '$, so that, from Eq.(28), the group and
phase velocities of de Broglie waves, as measured by A,
respectively are
 $$v'_g  = v' = \frac{{c_0 }}{{\left[ {1 + \left( {\frac{{\lambda
'}}{{\lambda _c }}} \right)^2 } \right]^{1/2} }} < c_0 ,\quad
\quad v'_p  = c_0 \left[ {1 + \left( {\frac{{\lambda '}}{{\lambda
_c }}} \right)^2 } \right]^{1/2}  > c_0            \eqno(37)$$ and
thus one finds that matter de Broglie waves perceive even the flat
space as a dispersive medium with an index of refraction
 $$\bar N_{flat}  = \left[ {1 + \left( {\frac{{\lambda '}}{{\lambda
_c }}} \right)^2 } \right]^{1/2}.                  \eqno(38)$$ One
recognizes that it is this $v_{g}'$ in Eq.(37), together with
$\frac{{\lambda '}}{{\lambda _c }} = \frac{{mc_0 }}{{p'}}$,
provides the energy transformation law:
 $$H = \frac{{mc_0^2 }}{{\sqrt {1 - \frac{{{v_{g}'}^2 }}{{c_0^2 }}} }}
                                                 \eqno(39)$$
Then, one recovers the special relativistic mass shell condition.
It follows that, in this case, the drag measured by B in terms of
his/her wavelength $\lambda$, is
 $$F_1^{flat}  = \frac{{\left( {\frac{\lambda }{{\lambda _c }}}
\right)^2 }}{{\left[ {1 + \left( {\frac{\lambda }{{\lambda _c }}}
\right)^2 } \right]^{} }} \times \left[ {1 - \frac{1}{{\left[ {1 +
\left( {\frac{\lambda }{{\lambda _c }}} \right)^2 } \right]^{1/2}
}}} \right].                                         \eqno(40)$$
As one can see, Eqs.(37)-(40) are restatements of the well known
special relativistic expressions, but only {\it interpreted} in a
different way.\\

\section{Fresnel Drag: Case 2}

According to first option (i), the Doppler shift is given by
Eq.(25). Thus, we have, to first order in $(V/c_0 )$:
 $$\lambda ' \approx \lambda (1 + V/Nc_0 ) = \lambda  + \Delta
\lambda ,\quad \quad\Delta \lambda = \frac{{\lambda V}}{{Nc_0}}.
                                         \eqno(41)$$
Then, writing again: $\bar N(r_0 ' = r_0 ,\,\lambda ') \equiv \bar
N(\lambda  + \Delta \lambda )$, we get from the Taylor expansion
 $$\bar N(\lambda  + \Delta \lambda ) \approx \bar N(\lambda ) +
\Delta \lambda \frac{{\partial \bar N}}{{\partial \lambda }} =
\bar N(\lambda )(1 + \frac{{\lambda V}}{{c_0 N\bar
N}}\frac{{\partial \bar N}}{{\partial \lambda }}). \eqno(42)$$
 The resultant group velocity as observed by B, who uses $\lambda$
of Eq.(41), is
 $$v_g  = \frac{{c_0 }}{{\bar N'(\lambda ')}}=\frac{{c_0 }}{{\bar
N(\lambda ')}} + \left( {1 - \frac{1}{{\bar N^2 (\lambda ')}}}
\right)V   $$
 $$\quad \quad \quad \quad  = \frac{{c_0 }}{{\bar N(\lambda+\Delta
\lambda )}} + \left( {1 - \frac{1}{{\bar N^2 (\lambda )}}}
\right)V = \frac{{c_0 }}{{\bar N(\lambda )}} + F_2 V,  \eqno(43)$$
where
 $$F_2  \equiv \left( {1 - \frac{1}{{\bar N^2 (\lambda )}}} \right) -
\frac{\lambda }{{N\bar N^2 (\lambda )}} \times \frac{{\partial
\bar N}}{{\partial \lambda }},                    \eqno(44)$$
 is the drag factor.  For light waves, $N=n$, $\bar N =n$  so that
 $$F_2  \equiv \left( {1 - \frac{1}{{n^2 (\lambda )}}} \right) -
\frac{\lambda }{{n^3 (\lambda )}} \times \frac{{\partial
n}}{{\partial \lambda }}.                   \eqno(45)$$
  This formula was first given by McCrea [23].
Writing explicitly, we find from Eq.(44),
 $$F_2  = 1 - \frac{1}{{n^2 (r_0 )}}.     \eqno(46)$$
This coefficient comes out to be independent of $\lambda$!
According to the second option (ii), B uses a vacuum wavelength.
In this case, the calculations would proceed slightly differently.
Consider Eq.(18) for $\omega'$ instead of Eq.(41). Then, we have,
to first order,
 $$\omega ' \approx \omega (1 - VN(\omega )/c_0 ) = \omega  + \Delta
\omega ,\quad \quad \Delta \omega  =  - \frac{{\omega VN(\omega
)}}{{c_0 }}.                                 \eqno(47)$$ Then,
proceeding as before,
 $$v_g (\omega ) = \frac{{c_0 }}{{\bar N(\omega )}} + \left[ {\left(
{1 - \frac{1}{{\bar N^2 (\omega )}}} \right) + \frac{\omega
}{{N(\omega )}} \times \frac{{\partial \bar N}}{{\partial \omega
}}} \right]V.                                   \eqno(48)$$ Now B
uses the vacuum wavelength to be $\lambda _0 = 2\pi c_0 /\omega$,
so that Eq.(48) gives
 $$v_g (\lambda _0 ) = \frac{{c_0 }}{{\bar N(\lambda _0 )}} + F_3 V,
                                              \eqno(49)$$
where
 $$F_3  = \left[ {\left( {1 - \frac{1}{{\bar N^2 (\lambda _0 )}}}
\right) - \frac{{\lambda _0 N(\lambda _0 )}}{{\bar N^2 (\lambda _0
)}} \times \frac{{\partial \bar N}}{{\partial \lambda _0 }}}
\right].                                   \eqno(50)$$
 For light waves, we get
 $$F_3  \equiv \left( {1 - \frac{1}{{n^2 (\lambda _0 )}}} \right) -
\frac{{\lambda _0 }}{{n(\lambda _0 )}} \times \frac{{\partial
n}}{{\partial \lambda _0 }}.                         \eqno(51)$$
This is the expression given by Lerche [1] and Cook {\it et al}
[2] for Fizeau experiment with water with index $n$. Writing
explicitly, we find from Eq.(50),
 $$F_3  = 1 - \frac{1}{{n^2 (r_0 )\left[ {1 + \left( {\frac{{\tilde
\lambda _0 }}{{\lambda _c }}} \right)^2 } \right]}} -
\frac{{\left( {\frac{{\tilde \lambda _0 }}{{\lambda _c }}}
\right)^2 }}{{\left[ {1 + \left( {\frac{{\tilde \lambda _0
}}{{\lambda _c }}} \right)^2 } \right]^2 }}           \eqno(52)$$
 where $\tilde \lambda _0  = \lambda _0 \Phi ^{ - 1}$. Thus,
so far, corresponding to  $N$ and $\bar N$, we have three Fresnel
coefficients $F_1$, $F_2$ and $F_3$ depending on the VAL and the
various Doppler shifted wavelengths used by B, as considered
in the literature.\\

\section{Fresnel drag: Case 3}

Consider an observer $\tilde A$ at rest with respect to the
gravitating source at a coordinate radial distance $r=r_{0}'$.
He/she will measure proper quantities. The mass shell condition
would be given by
 $$\hbar ^2 \tilde \omega '^2  = m^2 c_0^4  + c_0^2 \hbar ^2 \tilde
k'^2 ,\quad \quad \tilde k' = \Phi k',          \eqno(53)$$
 so that $\tilde A$ measures, in his neighborhood, the proper
phase and group velocities of the de Broglie waves which are
connected by
 $$\tilde v'_p \tilde v'_g  = \frac{{\tilde \omega '}}{{\tilde
k'}}\frac{{d\tilde \omega '}}{{d\tilde k'}} = c_0^2,
                                                  \eqno(54)$$
where, using Eq.(27),
 $$\tilde v'_p  = \frac{{\tilde \omega '}}{{\tilde k'}} = n\left(
{\frac{{\omega '}}{{k'}}} \right) = c_0 \tilde N(r'_0 ,\tilde
\lambda ') > c_0 ,\quad \tilde v'_g  = \frac{{d\tilde \omega
'}}{{d\tilde k'}} = \frac{{c_0 }}{{\tilde N(r'_0 ,\tilde \lambda
')}} < c_0 ,\quad \tilde N \equiv \frac{n}{N} > 1.  \eqno(55)$$
Note that, these phase and group velocities are not the same as
those measured by A, viz., Eqs.(27) and (28), which highlight the
difference between the two observers. The observer $\tilde A$
measures the velocity of light as ${\tilde{v}_{p}}'=
{\tilde{v}_{g}}'= c_0$ since $N=n$. Consider another observer
$\tilde B$ falling freely in the same radial direction attaining
an instantaneous speed $\tilde V$ at $r=r_{0}'$. Since the frame
in which $\tilde B$ is at rest is locally inertial in virtue of
the principle of equivalence, the speed of light measured by him
will also be $c_0$ and hence  $\tilde A$ and $\tilde B$ can be
connected by a Lorentz transformation. Then $\tilde{v}_{g}'$ would
appear to $\tilde B$  at $r=r_0$ as $\tilde{v}_{g}$given by the
VAL:
 $$\tilde v_g  = \frac{{\tilde v'_g  + \tilde V}}{{1 + \frac{{\tilde
V\tilde v'_g }}{{c_0^2 }}}}.                    \eqno(56)$$
Employing arguments similar to those in Cases 1 and 2, we can
straightaway write down the corresponding drag coefficients:

(a) $\tilde A$ measures $\tilde{\lambda}'$  and  $\tilde B$ uses
$\tilde{\lambda}$ connected by $\tilde \lambda ' \approx \tilde
\lambda (1 + \tilde V/c_0 )$:
$$\begin{array}{l}\tilde F_1  \equiv \left( {1 - \frac{1}{{\tilde N^2
(\tilde \lambda )}}} \right) - \frac{{\tilde \lambda }}{{\tilde N^2
 (\tilde \lambda )}} \times \frac{{\partial \tilde N}}{{\partial
 \tilde \lambda }} \\
 \quad  = \frac{{\left( {\frac{{\tilde \lambda }}{{\lambda _c }}}
 \right)^2 }}{{\left[ {1 + \left( {\frac{{\tilde \lambda }}
 {{\lambda _c }}} \right)^2 } \right]^{} }} \times \left[ {1 -
 \frac{1}{{\left[ {1 + \left( {\frac{{\tilde \lambda }}{{\lambda _c
 }}} \right)^2 } \right]^{1/2} }}} \right] \\
 \end{array}.                                  \eqno(57)$$

(b) $\tilde A$  measures $\tilde{\lambda}'$  and  $\tilde B$ uses
$\tilde{\lambda}$ connected by $\tilde \lambda ' \approx \tilde
\lambda (1 + \tilde V\tilde N/c_0 )$:
 $$\tilde F_2  \equiv \left( {1 - \frac{1}{{\tilde N^2 (\tilde
\lambda )}}} \right) - \frac{{\tilde \lambda }}{{\tilde N(\tilde
\lambda )}} \times \frac{{\partial \tilde N}}{{\partial \tilde
\lambda }} = 0.                                 \eqno(58)$$

 (c) $\tilde A$  measures $\tilde{\omega}$  and  $\tilde B$ uses
$\tilde{\lambda}_0$ connected by $\tilde \lambda _0  = 2\pi c_0
/\tilde \omega$:
 $$\begin{array}{l}
 \tilde F_3  \equiv \left( {1 - \frac{1}{{\tilde N^2 (\tilde
 \lambda _0 )}}} \right) - \frac{{\tilde \lambda _0 }}{{\tilde N^3
 (\tilde \lambda _0 )}} \times \frac{{\partial \tilde N}}{{\partial
  \tilde \lambda _0 }} \\
 \quad  = 1 - \frac{1}{{\left[ {1 + \left( {\frac{{\tilde \lambda _0
 }}{{\lambda _c }}} \right)^2 } \right]}} \times \left[ {1 +
 \frac{{\left( {\frac{{\tilde \lambda _0 }}{{\lambda _c }}} \right)^2
  }}{{\left[ {1 + \left( {\frac{{\tilde \lambda _0 }}{{\lambda _c }}}
   \right)^2 } \right]}}} \right], \\
 \end{array}                                         \eqno(59)$$
where $\tilde \lambda _0  = \lambda _0 \Phi ^{ - 1}$. We also see
that the radial proper velocity of the classical point particle as
measured by $\tilde A$  at $r=r_{0}'$  is given by
 $$\tilde v_{prop}'  = \frac{{dl'}}{{d\tau '}} =
n\frac{{dr'}}{{dt'}} = nv'_{coord}.                \eqno(60)$$
Using the definitions: $dl' = \Phi ^{ - 1} dr',\;d\tau ' = \Omega
dt',\;v'_{coord}  = \frac{{Nc_0 }}{{n^2 }}$, we find that $\tilde
v_{prop}'  = \tilde v'_g$. For light, of course, $v'_{coord}  =
\frac{{c_0 }}{n}$, and $\tilde v'_{prop}  = \tilde v'_g  = c_0$.
The last result is also consistent with the fact that $ds^2  =
c_0^2 d\tau '^2  - dl'^2  = 0$ gives $\frac{{dl'}}{{d\tau '}} =
c_0$. For light waves, we find $\tilde N =1$, so that $\tilde{F}_1
= \tilde{F}_2 =\tilde{F}_3 =0$. These indicate only the special
relativistic invariance of light speed, no matter what wavelength
$\tilde{B}$ uses. For de Broglie waves, the difference among the
drag coefficients is evident from Eqs.(57)-(59).\\

\section{Operational definitions}

In order to operationally realize the value of $F_1$  in a
gravitational field, consider a simple thought experiment. Let
there be a source in free space that produces de Broglie waves
with wavelength $\lambda'$. Then $\lambda$  is known via Eq.(31)
which is the wavelength measured by B. Let A take this source to
any point inside the refractive medium. Then, he will measure the
same $\lambda'$ as $\tilde \lambda ' = \lambda '\Phi ^{ - 1}$ and
B will find $\tilde \lambda  = \lambda \Phi ^{ - 1}$. The only
other quantity is the coordinate distance $r_0$ appearing in the
refractive index $n(r_{0})$ and $\Phi(r_{0})$. The expressions for
the index is supplied by the metric functions. For instance, in
the Reissner-Nordstrom field, with $G=c_{0}=1$, we have
 $$\Omega
^2 (r) = \left[ {1 - \frac{{(M^2  - Q^2 )}}{{4r^2 }}} \right]^2
\left[ {1 + \frac{M}{r} + \frac{{(M^2  - Q^2 )}}{{4r^2 }}}
\right]^{ - 2},                                   \eqno(61)$$
 $$\Phi ^{ - 2} (r) = \left[ {1 + \frac{M}{r} + \frac{{(M^2  - Q^2
)}}{{4r^2 }}} \right]^2,                             \eqno(62)$$
where $M$  and $Q$  are the mass and the electric charge. For the
Schwarzschild field, we have $Q=0$, so that
 $$n(r) = \frac{{\left( {1 + \frac{M}{{2r}}} \right)^3 }}{{\left( {1
- \frac{M}{{2r}}} \right)}}.                         \eqno(63)$$

If the relative velocity $V$  between A and B is small, $V<<c_0$,
we can take $r_0  \approx r_0 '$ from Eq.(29). If we consider that
both the observers are in a weak gravity field, we can take $r_0
\approx r_0 ' \approx l$, where $l$  is the physically measurable
distance from the center of the gravitating source to the field
point. Then
 $$n(r_0 ) \approx n(l) \approx 1 + \frac{{2M}}{l}.
                                             \eqno(64)$$
With these inputs, Eq.(36) provides the theoretically predicted
value of $F_1$  after the known value of the Compton wavelength is
plugged in.

Interesting results are obtained in the case of $F_2$  and
$\tilde{F}_2$. One finds that $F_2$  does not involve the
wavelength at all. This means that a Fizeau type experiment either
with light or with de Broglie waves would yield the same drag
factor, if Eq.(23) is followed in conjunction with Eq.(27). In
this case, it appears that the wavelength dependence introduced by
the group velocity is {\it undone} by Doppler shift. A similar
thing occurs also in the case of $\tilde{F}_2$ which is
identically zero.

\section{Comparison with real medium}

Starting from the wave equation in a nonuniformly moving fluid
with refractive index $n$, Leonhardt and Piwnicki [6] derive the
Lagrangian and the Hamiltonian for a light ray as observed by a
lab observer. From the action principle, they arrive at a
completely geometrical picture of ray optics in a moving medium.
Light rays are geodesic lines with respect to Gordon's metric,
which, in the comoving frame reads
 $$ds^2  = \frac{{c_0^2 }}{{n^2 }}dt'^2  - \left| {d\vec r'}
\right|^2.                                    \eqno(65)$$ The
Lagrangian, Eq.(49) of Ref. [6], they derived for a light particle
in the lab frame, is
 $$L =  - mc_0 \sqrt {c_0^2  - v^2  + \left( {\frac{1}{{n^2 }} - 1}
\right)\gamma ^2 \left( {c_0  - \frac{{\vec u.\vec v}}{{c_0 }}}
\right)^2 },                                        \eqno(66)$$
where $u$ is the fluid velocity in the lab frame,  $v\left( {
\equiv \frac{{v' + u}}{{1 + v'u/c_0^2 }}} \right)$ is the velocity
of the light particle conceived of having a fictitious mass $m$
and $\gamma ^2 = \left({1-\frac{{u^2 }}{{c_0^2 }}} \right)^{-1}$.
In the comoving frame of the fluid element, $\vec u =0$ so that
 $$L =  - mc_0^2  \times \frac{1}{n} \times \sqrt {1 - \frac{{v'^2
n^2 }}{{c_0^2 }}}.                                  \eqno(67)$$
Consider the Lagrangian for a massive particle in the comoving
frame, derived in our Ref.[18], viz.,
 $$L =  - mc_0^2 \Omega \left[ {1 - \frac{{v'^2 n^2 }}{{c_0^2 }}}
\right]^{1/2},                                    \eqno(68)$$
where $v'$  is the classical particle coordinate speed. Now note
that the metric (65) with $n$  as the real medium index, can be
obtained formally from Eq.(6) above simply by putting $\Phi =1$
and $\Omega =1/n$. Clearly, the $n$ in Eq.(68) has a {\it
different} origin: it derives from general relativity. Using this
value of $\Omega$ in Eq.(68), one finds that it is exactly the
same as Eq.(67).

The dispersion relation for light in the comoving frame $(\vec u
=0)$ following from Eq.(33) of Ref.[6] is
 $$\omega '^2  - c_0^2 k'^2  + (n^2  - 1)\omega '^2  = 0.
                                            \eqno(69)$$
This is precisely the same as that following from Eq.(15) with
$m=0$ for light.

Interestingly, taking a cue from Eq.(66), we may proceed to write
down the Lagrangian of the classical particle in the lab frame as
follows:

Our metric, Eq.(6) in the comoving frame can be written down as
 $$ds^2  = \Phi ^{ - 2}\times \left[ {c_0^2 dt'^2 - d\vec r'^2 +
\left( {\frac{1}{{n^2 }} - 1} \right)c_0^2 dt'^2 } \right].
                                                 \eqno(70)$$
To go to the lab frame, we effect a Lorentz transformation. Note
that there is a Lorentz invariant term in the parenthesis and
hence only the last term needs to be transformed. Thus, in the lab
frame the metric is
 $$ds^2  = g_{\mu \nu } dx^\mu  dx^\nu   = \Phi ^{ - 2}  \times
\left[ {\eta _{\mu \nu }  + \left( {\frac{1}{{n^2 }} - 1}
\right)V_\mu  V_\nu  } \right]dx^\mu  dx^\nu ,    \eqno(71)$$
where $\eta _{\mu \nu }  = \left[ {c_0^2 , - 1, - 1, - 1}
\right],\quad V_\mu   = \gamma \left( {1, - \frac{{\vec V}}{{c_0
}}} \right),\quad \gamma  = \left( {1 - \frac{{V^2 }}{{c_0^2 }}}
\right)^{ - 1/2}$, and  $\vec V$ is the velocity of our medium in
the lab frame. In the comoving frame, $V_{\mu}=(1,0,0,0)$. The
action is given by
 $$S =  - mc_0 \int {\sqrt {g_{\mu \nu } \frac{{dx^\mu
}}{{dt}}\frac{{dx^\nu  }}{{dt}}} } dt = \int {Ldt}$$ Defining
$v^\mu =\frac{{dx^\mu }}{{dt}}= \left( {1,\vec v} \right)$, we can
find the Lagrangian for a particle in the lab frame,
 $$L =  - mc_0 \Phi ^{ - 1}  \times \sqrt {c_0^2  - v^2  + \left(
{\frac{1}{{n^2 }} - 1} \right)\gamma ^2 \left( {c_0  - \frac{{\vec
V.\vec v}}{{c_0 }}} \right)^2 },                    \eqno(72)$$
The dispersion relation (or, the Hamiltonian) in the lab frame can
also be obtained by a Lorentz transformation on the mass shell
equation (15) in the comoving frame, rewritten as
 $$\omega '^2  - c_0^2 k'^2  + (n^2  - 1)\omega '^2  = \frac{{m^2
c_0^4 n^2 \Omega ^2 }}{{\hbar ^2 }}.                 \eqno(73)$$
Note that the right hand side is a Lorentz scalar and the left
hand side has a Lorentz invariant part $\omega '^2 - c_0^2 k'^2$.
The remaining part can be transformed to give
 $$\omega ^2  - c_0^2 k^2  + (n^2  - 1)\gamma ^2  \times \left(
{\omega  - \vec k.\vec V} \right)^2  = \frac{{m^2 c_0^4 n^2 \Omega
^2 }}{{\hbar ^2 }},                                   \eqno(74)$$
where $k_\mu =\left( {\frac{\omega }{{c_0 }}, -\vec k} \right)$ is
the wave four vector. There are several other ways in which
Eq.(74) could be obtained, either by usual Legendre
transformations from Eq.(72) or by the Hamilton-Jacobi equation
$g^{\mu \nu } \frac{{\partial S}}{{\partial x^\mu
}}\frac{{\partial S}}{{\partial x^\nu  }} = m^2 c_0^4$ with
$g^{\mu \nu }  = \Phi ^2  \times \left[ {\eta ^{\mu \nu }  +
\left( {n^2  - 1} \right)V^\mu  V^\nu  } \right]$. We do not do it
here.

A further interesting result holds as a corollary to Sec. 4: For
light waves in flat space, $v_p ' = v_g ' = c_0  = v_g$, as
expected. It should be noted that the Minkowski observers A and B
can also be located in the asymptotic region and the entire
analysis would remain the same. From the asymptotic vantage point,
these observers can see that, near the horizon, $n\to\infty$, then
$v_{p}',v_{g}' \to 0$, {\it both} for light and matter de Broglie
waves. It is exactly here that we find that the conditions for
optical black holes required by Leonhardt and Piwnicki [19] and
Hau {\it et al} [20] are provided {\it most naturally}, that is,
extremely low group velocity or high refractive index. In this
respect, optical and gravitational black holes look indeed
similar. Also, $v_{g}=V$ , implying that, while A sees everything
standstill at the horizon, B sees them moving away at the speed
$V$ because of his own relative motion. This is what we should
really expect.\\

\section{Summary and concluding remarks}

The present investigation is inspired by recent discoveries and
analyses of light propagation in Bose-Einstein condensates
[19,20]. The extremely low velocity of light in such condensates
lead to the possibility of creating optical analogs of
astrophysical black holes in the laboratory. In order to
theoretically model this possibility, Leonhardt and Piwnicki [6]
proceed from the moving optical medium to an effective gravity
field with a scalar curvature $R \ne 0$ in which light propagation
is shown to mimic that around a vortex core or optical black hole.
Novello and Salim [24] have recently shown that the propagation of
photons in a {\it nonlinear} dielectric medium can also be
described as a motion in an effective spacetime geometry. Our
approach here has been in the exact reverse direction: We proceed
from the gravity field and arrive at an effective optical
refractive medium and examine the theoretical consequences. The
motion of this medium is caused by the relative motion between the
observer B and the gravitating source.

We must mention that works based on the above mentioned analogies
provide some curious theoretical insights both in the real media
and in the gravitational field, as a result of wisdom borrowed
from one field and implanted into the other. This has been the
basic philosophy of the present paper. Many more interesting
results are known apart from the possibility of optical black
holes stated above. For instance, an analysis in acoustic theory
leads to a remarkable result that the Hawking radiation in black
hole physics is not of dynamical, but {\it kinematical} origin
(Visser, Ref.[9]). Conversely, a gravitational refractive index
approach, similar in spirit to that of ours, has yielded the
possibility of {\it \v{C}erenkov radiation} in the outskirts of a
wormhole throat [26-28]. In the present paper, we envisaged a
nontrivial {\it dispersive Fresnel drag} coefficient in a gravity
field. We must emphasize that these results are only of pedagogic
interest at present. A further confirmation or otherwise of these
results would establish the extent to which these analogies could
actually be relied upon.

We saw above how dispersion effects, both for massless and massive
particles, appear naturally as a consequence of the systematic
development of an effective medium approach to gravitational
field. Various expressions for the drag coefficients result due to
the use of VAL and different wavelengths used by the observer B.
(See Refs.[1-3] for more detailed arguments on the question of the
use of appropriate wavelength). It is demonstrated that $F_2$  is
{\it independent} of $\lambda$ even in a dispersive medium for
massive particles and that $\tilde{F}_2$ is identically zero.
These results may have interesting implications for both optical
and general relativity black holes.

It does not seem easy to simulate real experiments, with our type
of unbounded medium, that parallel those dealing with ordinary
media like solid, fluid or superfluid. For this reason, we limited
ourselves only to theoretical calculations of the drag
coefficients and the expressions may be useful in the study of
passage of light and cosmic particles in astrophysical media since
what we actually see from the moving Earth is not what was
originally sent from the source. This work is underway.

We saw that the present analysis naturally complements the curved
space analogy of a moving medium. Some of the key expressions in
the comoving frame are indeed the same. Moreover, we can find a
direct extension of the expressions to a genuine gravity field
(Sec. 8). The resulting Lagrangian and Hamiltonian describe the
trajectories of a particle as viewed from the lab frame, say, a
rocket. It also appears that the nomenclature ``optical black
hole" is quite apt as the conditions required for their creation
are most naturally met near the gravitational horizon. This gives
an indication that the behavior of the real optical medium should
mimic that of our equivalent refractive medium around a coordinate
singularity. A favorable situation is attained if light perceives
the highly refractive real optical medium as dispersionless which,
in our effective medium, is actually the case. Leonhardt and
Piwnicki [6] also make a similar statement in the context of their
vortex analysis. It is interesting to note that an index of the
form  $n=C/r$, where $C$ is a constant, when put in Eq.(4) yields
orbits that resemble those around an optical vortex core [10]. A
similar investigation with a different form of index  has been
reported also recently [29].\\

{\bf Acknowledgments}

One of us (KKN) wishes to thank the Director, Professor Ouyang
Zhong Can, for providing hospitality and excellent working
conditions at ITP, CAS.  The work was in part supported by the
TWAS-UNESCO visiting associateship program of ICTP, Italy, and
also by NNSFC under Grant Nos. 10175070 and 10047004, as well as
by NKBRSF G19990754.


\begin{thebibliography}{99}
\bibitem{1}I. Lerche, Am. J. Phys. {\bf 45},1154 (1977).

\bibitem{2}R.J. Cook, H. Fearn, and P.W. Milonni, Am. J. Phys.
 {\bf 63}, 705 (1995).

\bibitem{3}J.L. Synge, Relativity: The Special Theory (North Holland,
  Amsterdam, 1965).

\bibitem{4}G.A. Sanders and S.Ezekiel, J. Opt. Soc. Am. B {\bf 5},
 674 (1988).

\bibitem{5}P. Zeeman, Proc. Roy. Soc. Amsterdam {\bf 17}, 445 (1914);
  Arch. Neerl. Sci. Exactes Nat. {\bf 3A}, 10, 131 (1927).
  These experiments, conducted over 14 years, have so far been
  regarded as
  authoritative and, as analyzed by Lerche [1], they lean slightly
  in favor of Lorentz's form, Eq.(45). However, he thoroughly
  scrutinizes the details of the experiment and finds that Zeeman's
  results are essentially inconclusive. He suggests a repetition of
  the experiment.

\bibitem{6} U. Leonhardt and P. Piwnicki, Phys. Rev. A {\bf 60},
   4301 (1999).

\bibitem{7}M.V. Berry, R.G. Chambers, M.D. Large, C. Upstill,
 and J. Walmsley, Eur. J. Phys. {\bf 1}, 154 (1980).

\bibitem{8}P. Roux, J. de Rosney, M. Tanter, and M. Fink,
  Phys. Rev. Lett. {\bf 79}, 3170 (1997).

\bibitem{9}H. Davidowitz and V. Steinberg, Europhys.
 Lett. {\bf 38}, 297 (1997);
 For Unruh's acoustic "dumbhole", see: W.G. Unruh,
 Phys. Rev. Lett. {\bf 46}, 1351 (1981); Phys. Rev.
 D {\bf 51}, 2827 (1995); see also: T.A. Jacobson, Phys. Rev.
 D {\bf 44}, 1731 (1991); M. Visser, Class. Quant.Grav.15,1767
 (1998), Phys. Rev. Lett. {\bf 80}, 3436 (1998); {\it ibid}, {\bf 85},
  5252 (2000) and other references cited in Ref.[1].

\bibitem{10} J. Evans and M. Rosenquist, Am. J. Phys. {\bf 54},
 876 (1986); J. Evans, Am. J. Phys. {\bf 58}, 773 (1990);
  M. Rosenquist and J. Evans, Am. J. Phys. {\bf 56}, 881 (1988);
   J. Evans, Am. J. Phys. {\bf 61}, 347 (1993).

\bibitem{11}K.K. Nandi and A. Islam, Am. J. Phys. {\bf 63},
  251 (1995).

\bibitem{12}A.S. Eddington, Space, Time and Gravitation
 (Cambridge U.P., Cambridge, 1920, reprinted 1987).

\bibitem{13}F. de Felice, Gen. Relat. Grav. {\bf 2}, 347 (1971);
 F.R.Tangherlini, Am. J. Phys. {\bf 36}, 1001 (1968);
  Phys. Rev. A {\bf 12}, 139 (1975); B. Bertotti, Gen. Rel. Grav.,
  {\bf 30}, 209 (1998).

\bibitem{14}W. Gordon, Ann. Phys. (Leipzig) {\bf 72}, 421 (1923);
 P.M. Quan, C.R. Acad. Sci. Paris, {\bf 242}, 465 (1956);
  Arch. Ration. Mech. Anal. {\bf 1}, 54 (1957/58);
  J. Plebanski, Phys. Rev. {\bf 118}, 1396 (1960).

\bibitem{15}J. Evans, K.K. Nandi, and A. Islam, Gen. Relat. Grav.
 {\bf 28}, 413 (1996).

\bibitem{16}J. Evans, K.K. Nandi and A. Islam, Am. J. Phys.
 {\bf 64}, 1404 (1996).

\bibitem{17}P.M. Alsing, Am. J. Phys. {\bf 66}, 779 (1998).

\bibitem{18}J. Evans, P.M. Alsing, S. Giorgetti, and K.K. Nandi,
 Am. J. Phys. {\bf 69}, 1103 (2001).

\bibitem{19}U. Leonhardt and P. Piwnicki, Phys. Rev. Lett.
  {\bf 84}, 822 (2000); {\it ibid}, {\bf 85}, 5253 (2000).

\bibitem{20}L.V. Hau, S.E. Harris, Z. Dutton, and C.H. Behroozi,
 Nature (London), {\bf 397}, 594 (1999).

\bibitem{21}L. Stodolsky, Gen. Relat. Grav. {\bf 11}, 391 (1979).

\bibitem{22}P.M. Alsing, J.C. Evans, and K.K. Nandi,
  Gen. Relat. Grav. {\bf 33}, 1459 (2001).

\bibitem{23}W.H. McCrea, Relativity Physics (Methuen, London,1952).

\bibitem{24}M. Novello and J.M. salim, Phys. Rev.
  D {\bf 63}, 083511 (2001).

\bibitem{25}See, for instance: K.K. Nandi, J.C. Evans and A. Islam,
  Int. J. Mod. Phys. A {\bf 12}, 3171 (1997).

\bibitem{26}L.A. Anchordoqui, S. Capozziello, G. Lambiase,
  and D.F. Torres, e-print gr-qc/0011097.

\bibitem{27}E.F. Beall, Phys. Rev. Lett. {\bf 21}, 1364 (1968);
  Phys. Rev. D {\bf 1}, 961 (1970).

\bibitem{28}A. Gupta, S. Mohanty, and M.K. Samal, Class. Quantum
 Grav. {\bf 16}, 291 (1999).

\bibitem{29}M. Marklund, D. Anderson, F. Cattani, M. Lisak
 and L. Lundgren, Am. J. Phys. {\bf 70}, 680 (2002).

\end{thebibliography}
\end{document}